\documentclass[a4paper,fleqn,usenatbib]{mnras}

\usepackage{newtxtext,newtxmath}
\usepackage[T1]{fontenc}
\usepackage{ae,aecompl}

\usepackage{graphicx}	% Including figure files
\usepackage{amsmath}	% Advanced maths commands
\usepackage{amssymb}	% Extra maths symbols

\usepackage{tabularx}
\usepackage{float}
\usepackage[utf8]{inputenc}
\usepackage[english]{babel}

\usepackage[usenames,dvipsnames]{xcolor}

\newcommand{\Pa} {P_\text{a}}
\newcommand{\Ps} {P_\text{s}}
\newcommand{\Pt} {P_\text{tot}}
\newcommand{\PaINI} {P_\text{a,0}}
\newcommand{\Pone} {P_\text{aa}}
\newcommand{\PoneINI} {P_\text{aa,0}}

\definecolor{owngreen}{rgb}{0,0.5,0}
\newcommand{\Pddpl}	{P_\text{MPL}}
\newcommand{\Ptddpl}	{P_\text{tMPL}}

\newcommand{\Pddplt}	{P_\text{MPL,t}}
\newcommand{\Pddplxi}	{P_\text{MPL,$\xi$}}

\title[MPL Distribution for IMF]{A Multiple Power Law Distribution
	for Initial Mass Functions}

\author[C.~Essex et al.]{
Christopher Essex,$^{1}$
Shantanu Basu,$^{2}$\thanks{E-mail: basu@uwo.ca}
Janett Prehl,$^{3}$
and Karl Heinz Hoffmann$^{3}$\thanks{E-mail: hoffmann@physik.tu-chemnitz.de}
\\
$^{1}$Department of Applied Mathematics, The University of Western Ontario, London, ON, N6A 5B7, Canada\\
$^{2}$Department of Physics \& Astronomy, The University of Western Ontario, London, ON, N6A 3K7, Canada\\
$^{3}$Institut für Physik, Technische Universität Chemnitz, D-09107 Chemnitz, Germany
}

\date{Accepted 2020 March 13. Received 2020 March 13; in original form 2019 October 7}

\pubyear{2020}

\begin{document}
\label{firstpage}
\pagerange{\pageref{firstpage}--\pageref{lastpage}}
\maketitle

\begin{abstract}
% \RED{padoan.p.04.mysterious.559
% should have Nordlun\emph{d}}
% \newline

We introduce a new multi-power-law distribution for the 
Initial Mass Function (IMF) 
to explore its potential properties. 
It follows on prior work that introduced mechanisms 
accounting for mass accretion in star formation, 
developed within the framework of general evolution equations for the mass distribution
of accreting and non-accreting (proto)stars. 
This paper uses the same fundamental framework to demonstrate
that the interplay between a mass-dependent and a time-dependent 
step-like dropout rate from accretion leads to IMFs
that exhibit multiple power laws
for an exponential mass growth.
While the mass-dependent accretion and its dropout is intrinsic to each star,
the time-dependent dropout might be 
tied to a specific history 
such as the rapid consumption of nebular material by nearby stars 
or the sweeping away of some material by shock waves. 
The time-dependent dropout folded into the mass-dependent
process of star formation is shown to have a significant influence on the IMFs. 

\end{abstract}

\begin{keywords}
accretion -- stars: formation -- stars: initial mass function
\end{keywords}

%%%%%%%%%%%%%%%%%%%%%%%%%%%%%%%%%%%%%%%%%%%%%%%%%%

%%%%%%%%%%%%%%%%% BODY OF PAPER %%%%%%%%%%%%%%%%%%

\section{Introduction}

The distribution of stellar (and substellar) masses resulting from the star-formation 
-- known as the initial mass function (IMF) -- 
provides the initial conditions for subsequent main-sequence or
brown-dwarf evolution. 
There are a variety of ideas for the mechanisms leading to the IMF.
One idea rests on turbulence in molecular clouds 
which determines the core mass function (CMF).
This then maps directly onto the 
IMF 
\citep{padoan.p.02.stellar.870,%
padoan.p.04.mysterious.559,%
hennebelle.p.08.analytical.395,%
hennebelle.p.09.analytical.1428}
in the sense that approximately a fixed proportion of the core mass 
goes into the star(s).
The IMF has been treated very generally as a statistical outcome of multiple
contributing phenomena, with random multiplicative processes yielding a 
lognormal distribution 
\citep{zinnecker.h.84.star.43}. 
Another idea is that 
accretion processes set the intermediate and high mass power law tail of 
the IMF. 
\cite{zinnecker.h.82.prediction.226} 
showed that a Bondi accretion process 
would broaden an initial distribution of masses and lead to a power
law tail that was similar to that observed for the IMF. 
\cite{adams.f.96.theory.256} 
advanced the idea that the observed IMF 
is dominated more by the accretion termination processes than by the CMF.
These latter two approaches put more emphasis on the temporal evolution of the actual
star formation process.

Some of these
ideas were combined in a hybrid model for the IMF, 
the 
modified lognormal power-law (MLP) model
\citep{basu.s.04.power-law.L47,%
basu.s.15.mlp.2413}
that relied on the statistical picture for the initial seed masses 
and a deterministic exponential growth law for masses coupled 
with a probabilistic model for the termination times of the growth. 
The initial seed masses were taken to 
follow a lognormal distribution, but each object then accreted mass in a
deterministic (exponential growth) fashion. The termination times of accretion
were then chosen statistically from an exponential distribution. This is 
consistent with equally likely stopping probability of accretion in equal
time intervals. 
However, 
\cite{basu.s.04.power-law.L47} 
also pointed out that the exponential
distribution of termination times was consistent with the deterministic
model of an exponentially growing rate of seed production up until some
final time $T$ when star-formation ends. In the MLP model, 
the limit $T \rightarrow \infty$ is employed, although this assumption can be relaxed. 
Furthermore, it is worth noting that the MLP distribution becomes a pure
power-law distribution in the limit that the initial lognormal seed 
distribution has zero variance, i.e., is a delta function. The index of the
power law in the differential number per logarithmic mass bin is
$\alpha \equiv \delta/\gamma$, 
which is the dimensionless ratio of the 
termination rate $\delta$ of accretion and the growth rate $\gamma$ of mass 
in an individual object.

\cite{myers.p.11.star.98} 
developed a model for the IMF based on equally likely stopping 
and also a hybrid mass growth law that had an initially constant mass accretion rate 
that later transitioned (at a specific mass) into a rapidly increasing value 
that could be characterized as exponential growth, Bondi accretion, 
or some intermediate growth law. 
He found a peaked IMF 
even when starting with all protostars having zero mass at an initial time. 
This led to an IMF that resembled the field star IMFs of 
\cite{kroupa.p.02.initial.82} 
and 
\cite{chabrier.g.05.initial.41} 
for suitable values of the transition mass, the characteristic timescale of accretion termination, and parameters describing the rapid mass growth at late times. 

Recently, \cite{hoffmann.k.18.dual.2113} showed that it is possible to 
produce a peaked IMF with dual power law (DPL) behavior when all
protostars start with very small initial masses, thereby also 
removing the need to start with an an initial lognormal distribution of 
seed masses to produce a peaked IMF.
The derivation in the following sections is
a direct sequel of that paper.
We advise the reader to consult \cite{hoffmann.k.18.dual.2113} 
in order to develop a deeper appreciation of the formalism. 
\cite{hoffmann.k.18.dual.2113} developed the peaked IMF 
based on a special case of general evolution equations
for the distribution function of protostars still accreting and that for stars 
that are no longer accreting. 
These evolution equations account for the rearrangement in time 
of the probability to find a (proto)star of a certain mass
due to the accretion of mass or other specifiable processes
that need not be purely deterministic or stochastic in nature. 
The evolution equations in play are sufficiently general that,
for example, the MLP can be rediscovered through these means. 
The new feature in
\cite{hoffmann.k.18.dual.2113} 
is a dropout rate 
that is initially small but rises to its final value in a sigmoid fashion
at a certain time (and therefore also mass)
and with a characteristic temporal width $\eta^{-1}$. 
``Sigmoid" means that the function
transits in an s-shaped form from its $-\infty$-limit
to its $\infty$-limit.
One can envision this transition as a smoothed
or smeared out Heavyside step function.
Accordingly, for greater masses there is an increased probability that a star will stop accretion.
In that sense the accretion in stars is a self-limiting process. 
The result is a dual power law distribution (DPL).  
The DPL has a rising
power law $\text{d}n/\text{d}\log M \propto M^{2\beta}$ 
at low mass with index $\beta=\eta/\gamma$, 
in which $\eta$ is the dropout rise rate, and a declining power law 
at intermediate and high mass $\text{d}n/\text{d}\log M \propto M^{-\alpha}$ 
with index $\alpha = \delta/\gamma$. 
The peak of the distribution is approximately at $m_{\rm S}$, 
the mass at which the transition to greater termination rate takes place. 
This transition may be physically associated 
with the onset of nuclear fusion within the protostar. 
This agrees with the generation of outflows at the onset of nuclear fusion, 
as suggested by 
\cite{shu.f.87.star.23}. 
Interestingly, such an IMF with two power laws has been suggested 
by observations of the ONC by
\cite{rio.n.12.initial.14}
and of the 25 Orionis group by
\cite{suarez.g.19.system.05739v1}.

The DPL model is a direct consequence 
of the increased dropout rate from stellar accretion beyond a certain mass.
By its nature, it exhibits intrinsic properties of the star formation process
and thus would induce universal properties  of the type sought for the IMF. 
But accretion need not only be terminated due to intrinsic properties. 
Extrinsic  growth limitation is also permissible in the evolution equation framework. 
This type of growth limitation has a distinct and different significance. 
While intrinsic limitation is associated with properties of a single star's growth, 
extrinsic limitation may be associated with particular histories of the larger system.

Termination of accretion can be attributed to many possible events. 
These include dynamical ejection of protostars from small multiple systems,
outflow driven clearing of surrounding gas, 
a finite available mass reservoir due to limits in core mass 
or a competition with neighboring accretors, 
and stellar feedback from massive stars 
in the same cluster associated with stellar winds, 
outflows, 
or ionization fronts 
(see review by \cite{bonnell.i.07.origin.149}
and references therein for a discussion of these mechansims). 
These processes may follow a temporal order 
that is approximately the order written above. 
The ejections typically occur at the earliest times, 
outflows may require the onset of at least deuterium fusion, 
and the effect from nearby massive stars may be felt much later if at all.
It is therefore reasonable to explore the idea of a further increase of
the termination probability at a later time than the initial rise, 
which can be loosely called a ``nebula time''. So for example, the initial rise
of termination probability may reflect the transition from dynamical ejection driven
termination to outflow driven termination, while the second rise may represent the
transition from outflow driven termination to a mass starvation driven termination.
The mass starvation could of course occur due to multiple causes as listed above. However,
we resist the impulse to add even more steps of an increase in termination probability,
focusing instead on the effect of a single notable change to the generative
model for the DPL.

To illustrate the consequences of a second increase of the dropout rate, 
we return to the evolution equation machinery.
This leads to a variation on the DPL model 
that introduces a second power law at intermediate to high masses,
apart from the high mass power law decrease of the IMF.
Interestingly, a three power law IMF is shown by 
\cite{kroupa.p.01.variation.231,%
kroupa.p.02.initial.82}.
We follow the same methodology as in 
\cite{hoffmann.k.18.dual.2113}, 
employing an evolutionary approach that 
accounts for both accreting objects and those that have dropped out of the 
accretion process according to a specified dropout rate. 
This methodology is quite general and can be expanded 
to increasing levels of complexity in future work. 
It also emphasizes the time-dependent nature of the mass function
within any star-forming region. 
Recent work by 
\cite{drass.h.16.bimodal.1734} 
and
\cite{jerabkova.t.19.when.06974v2} 
shows 
that the Orion Nebula Cluster has had several episodes of star formation 
evidenced by distinct stellar populations of different ages. 

Section 2 of this paper presents the derivation of the new multi-power-law 
distribution, while Section 3 illustrates its various properties and limits.
Section~4 shows a comparison with some published IMF estimates 
and our conclusions are stated in Section~5.

%SECTION%%SECTION%%SECTION%%SECTION%%SECTION%%SECTION%%SECTION%
%SECTION%%SECTION%%SECTION%%SECTION%%SECTION%%SECTION%%SECTION%
\section{A Probabilistic Description of Star Formation}

The IMF describes the probability to find a star of mass $M$ in a star cluster.
Here its probability density function is denoted as $P_\text{IMF}$.
It will prove convenient to work in terms of $m=M/M_\odot$ 
or a monotonic increasing function $\xi$ of mass $m$, 
(e.g. $\xi=\log_{10} m$ or $\xi=\ln m$) to be specified as needed. 
We will none the less refer to $\xi$ as ``mass".

%SUBSECTION%%SUBSECTION%%SUBSECTION%%SUBSECTION%%SUBSECTION%
\subsection{The Accretion-Dropout Rates}

A core element in our development is the dropout rate, 
which determines that fraction of stars which stop accretion
as a function of time and mass. 
That fraction then becomes part of the IMF.
The overall dropout rate we choose must reflect 
the two different sources of accretion stopping discussed above:
the intrinsic and the extrinsic ones.
Here we will thus employ an accretion dropout rate of special additive structure,
namely 
\begin{align}
k(\xi,t) = k_\text{$\xi$}(\xi) + k_\text{t}(t),
\label{eq-kt+kx-Def}
\end{align}
where the mass-related term $k_\text{$\xi$}(\xi)$ 
reflects the intrinsic accretion stopping processes
while the time-related term $k_\text{t}(t)$ 
reflects the extrinsic accretion stopping processes.
As these processes are envisioned as occurring independent
from each other, 
the corresponding accretion dropout rates should simply add.
Each of those 
are chosen as sigmoid functions in the same fashion
as in our previous work:
\begin{align}
	k_\text{t}(t) &= \kappa_\text{t}/2 (1+\tanh[(t-t_\text{St})\lambda_\text{t}]  ),
\label{eq-kt-Def} 
\\
	k_\text{$\xi$}(\xi) &= \kappa_\text{$\xi$}/2 (1+\tanh[(\xi-\xi_\text{S$\xi$})\beta_\text{$\xi$}]  ).
\label{eq-kxi-Def} 
\end{align}
Both $t$ and $\xi$ have domains $-\infty$ to $\infty$. 

Here $t_\text{St}$ is the characteristic (switch) time 
when the time-dependent dropout processes are switched on and 
lead to a maximum dropout rate of $\kappa_\text{t}$,
which is the asymptotic value of the sigmoid function.
$\xi_\text{S$\xi$}$ is the characteristic (switch) mass 
where the mass-dependent dropout processes are switched on
and result in a maximum dropout rate of $\kappa_\text{$\xi$}$,
which is the corresponding value for the mass case.
The time and mass span of the transition zones are characterized by $1/\lambda_\text{t}$
and 
$1/\beta_\text{$\xi$}$ respectively.

%SUBSECTION%%SUBSECTION%%SUBSECTION%%SUBSECTION%%SUBSECTION%
\subsection{Model Structure}

The basis for our IMF theory is the overall probability density $\Pt(\xi,t)$ 
to find a condensation of mass $\xi$ at time $t$ in a cluster.
Its normalization obeys $ \int  \Pt(\xi,t) \text{d}\xi= 1 $,
which is reflected in the notation $ \Pt(\xi,t) \,\text{d}\xi $ 
if we see the necessity to uniquely describe the mass variable 
with respect to which $ \Pt(\xi,t) $ is a probability density.
$\Pt(\xi,t)$ is the sum of the two probabilities $\Pa(\xi,t)$ and $\Ps(\xi,t)$,
where $\Pa(\xi,t)$ is the probability to find a condensation of mass $\xi$ still 
actively
accreting at time $t$, 
and $\Ps(\xi,t)$ is the probability to find a condensation of mass $\xi$ 
that has 
stopped
accreting
(and has thus become a member of the IMF):
\begin{align}
\Pt(\xi,t) = \Pa(\xi,t) + \Ps(\xi,t). 
\label{eq-P}
\end{align}
The time evolution of $\Pa(\xi,t)$ is described by
\begin{align}
\partial_t \Pa(\xi,t) = {\cal L}  \Pa(\xi,t) - k(\xi,t) \Pa(\xi,t),
\label{eq-Pa-Dyn}
\end{align}
where ${\cal L}$ is an operator describing the changes in $\Pa(\xi,t)$ 
due to the accretion process, 
and $ k(\xi,t) \Pa(\xi,t) $ describes the changes in $\Pa(\xi,t)$ 
due to the dropout processes. 
Thus $k(\xi,t) \Pa(\xi,t)$ is that part of $\Pa(\xi,t)$,
which becomes inactive, or stationary, per time unit
and thus a member of the IMF.
Note, that the evolution operator ${\cal L}$ might describe 
a deterministic accretion of mass,
but it might also be a probabilistic operator like a Fokker-Planck operator. 

The evolution equation (\ref{eq-Pa-Dyn}) is supplemented with
an initial condition $\Pa(\xi,t_0) = \PaINI(\xi)$
which we choose, as in our previous paper,
to be 
\begin{align}
\PaINI(\xi)
= 
\hat{\delta}(\xi - \xi_0) ,
\label{eq-PaINI-DeltaFunc}
\end{align}
where $\hat{\delta}$ represents the Dirac delta function, 
in contrast to the unrelated constant $\delta$ used in the introduction. 
This choice reflects that all condensations are assumed 
to start with the same mass $\xi_0$.

As $\Ps(\xi,t)$ is the probability to find a condensation of mass $\xi$ 
which has stopped accreting,
it can be calculated 
by the time integral of all deactivated parts of $\Pa(\xi,t)$
and
thus obeys
\begin{align}
\Ps(\xi,t) = \int_{t_0}^{t}  k(\xi,t') \Pa(\xi,t') \text{d}t',
\label{eq-Ps-Dyn}
\end{align}
with $t_0$ being the 
time for the onset of star formation,
which without loss of generality 
is set to zero.
The IMF is then given by
\begin{align}
P_\text{IMF}(\xi) = \lim_{t \to \infty} \Ps(\xi,t) = \int_{0}^{\infty}  k(\xi,t') \Pa(\xi,t') \text{d}t'.
\label{eq-PIMF-Dyn}
\end{align}

The limit $t \to \infty$ is a simplification in the following sense:
In the statistical description of distributions the Gaussian distribution,
which has an infinite domain,
is often used for a variable 
which has a finite domain.
For instance, the height of people have a finite range,
none the less the Gaussian is used for its description.
This is effective, since there is very little probability 
in the tails of the distribution,
but then -- in order to keep the normalization correct --
one has to extend the integration to infinity.
While the star formation in a cluster is occurring 
over a finite time and with a finite mass span,
we will for simplicity sometimes extend the time and mass regime 
to $-\infty$ to $\infty$.

%SUBSECTION%%SUBSECTION%%SUBSECTION%%SUBSECTION%%SUBSECTION%
\subsection{Accretion Rate}

Following the literature
\citep{basu.s.04.power-law.L47,%
basu.s.15.mlp.2413,%
hoffmann.k.18.dual.2113}
we now make a further simplifying assumption, 
namely that the mass accretion is proportional to the already accreted mass.
In the literature several different accretion growth laws have been proposed,
for example the Bondi rate has mass accretion proportional to mass squared.
Here we will 
restrict ourselves 
to a deterministic evolution law
with a mass accretion proportional to the already accreted mass
\begin{align}
\frac{\text{d} \ln m}{\text{d}t} = \gamma.
\label{eq-LNmDot}
\end{align}
Thus a condensation with initial mass $m_0$ at time $t_0=0$ 
is transported in time $t$ to $m = m_0 e^{\gamma t}$.
We now introduce the probability distribution $\Pone(\xi,t)$,
which would evolve from the initial distribution $\Pone(\xi,0)=\PaINI(\xi)$
\emph{for a vanishing dropout rate}, i.e. $k(\xi,t)=0$,
\begin{align}
\partial_t \Pone(\xi,t) = {\cal L}  \Pone(\xi,t) ,
\label{eq-Pr-Dyn}
\end{align}
and which keeps its normalization as all condensations are always active.
With the choice $\xi=\ln m$ 
the corresponding evolution operator is then 
$ {\cal L}[\xi] = - \gamma \partial_{\xi}$
and we find
\begin{align}
\partial_t \Pone(\xi,t) 
=
 - \gamma \partial_{\xi} \Pone(\xi,t).
\label{eq-Pone-MLP-Dyn}
\end{align}
This is simply the half wave equation, which then shifts the initial 
distribution of $\Pone(\xi,0)$ uniformly to greater masses, 
\begin{align} 
\Pone(\xi,t) 
&= \PoneINI(\xi - \gamma t)
= \hat{\delta}(\xi- \xi_0 - \gamma t),
\label{eq-Pone-MLP-Dyn2}
\end{align}
where eq.~(\ref{eq-PaINI-DeltaFunc}) has been used.

%SUBSECTION%%SUBSECTION%%SUBSECTION%%SUBSECTION%%SUBSECTION%
\subsection{Accretion Dropout Rate}

Now, we cross over from a vanishing accretion dropout rate 
($k(\xi,t)=0$) to
a finite accretion dropout rate ($k(\xi,t)\neq0$).
Let us assume that the evolution operator only depends on $\xi$,
${\cal L} = {\cal L}[\xi]$,
and
the accretion-dropout rate is the sum of two contributions,
as given in eq.~(\ref{eq-kt+kx-Def})
\begin{align}
k(\xi,t) = k_\text{$\xi$}(\xi) + k_\text{t}(t) , 
\nonumber
\end{align}
Then (\ref{eq-Pa-Dyn}) becomes
\begin{align}
\partial_t \Pa(\xi,t) ={\cal L}[\xi] \Pa(\xi,t) - (k_\text{$\xi$}(\xi) + k_\text{t}(t)  ) \Pa(\xi,t).
\label{eq-Pa-MLP-Dyn}
\end{align}
We then use 
\begin{align}
\Pa(\xi,t) = g_\text{$\xi$}(\xi) g_\text{t}(t) \Pone(\xi,t),
\label{eq-Pa-ansatz}
\end{align}
as an ansatz to decompose eq. (\ref{eq-Pa-MLP-Dyn}), 
\begin{align}
g_\text{$\xi$}(\xi)& \Pone(\xi,t)  \; \partial_t g_\text{t}(t) 
+ g_\text{$\xi$}(\xi) g_\text{t}(t) \; \partial_t \Pone(\xi,t)
\nonumber
\\ 
=&
g_\text{t}(t) (- \gamma \partial_{\xi} g_\text{$\xi$}(\xi) \Pone(\xi,t)) 
\nonumber
\\
&- (k_\text{$\xi$}(\xi) + k_\text{t}(t)  )  g_\text{$\xi$}(\xi) g_\text{t}(t) \Pone(\xi,t)
\nonumber
\\ 
=&
- \gamma g_\text{t}(t) g_\text{$\xi$}(\xi) \partial_{\xi} \Pone(\xi,t)
- \gamma g_\text{t}(t) \Pone(\xi,t) \partial_{\xi} g_\text{$\xi$}(\xi)  
\nonumber
\\
&- k_\text{$\xi$}(\xi) g_\text{$\xi$}(\xi) g_\text{t}(t) \Pone(\xi,t)
- k_\text{t}(t)  g_\text{$\xi$}(\xi) g_\text{t}(t) \Pone(\xi,t).
\label{eq-Pa-MLP-Dyn-1}
\end{align}
Dividing eq.~(\ref{eq-Pa-MLP-Dyn-1}) 
by our product ansatz for $\Pa(\xi,t)$
we get 
\begin{align}
&\frac{\partial_t g_\text{t}(t)}{g_\text{t}(t)} + k_\text{t}(t) 
+ \frac{\gamma \; \partial_\xi g_\text{$\xi$}(\xi)}{g_\text{$\xi$}(\xi)} 
	+ k_\text{$\xi$}(\xi)
\nonumber
\\ 
&+ \frac{\partial_t \Pone(\xi,t)}{\Pone(\xi,t)} 
+ \frac{\gamma \; \partial_\xi \Pone(\xi,t)}{\Pone(\xi,t)}
=0
\label{eq-Pa-Decomp-Help}
\end{align}
Using eq.~(\ref{eq-Pone-MLP-Dyn}) we see that the last two terms
in the above equation cancel and we are left with
\begin{align}
\left(  \frac{\partial_t g_\text{t}(t)}{g_\text{t}(t)} + k_\text{t}(t)  \right)
+
\left(  \frac{\gamma \; \partial_\xi g_\text{$\xi$}(\xi)}{g_\text{$\xi$}(\xi)} + k_\text{$\xi$}(\xi)  \right)
=0.
\label{eq-Pa-Decomp-Help-2}
\end{align}
Generalizing the approach we have taken in our previous paper 
\citep{hoffmann.k.18.dual.2113},
we split the above equation into two parts
to obtain separate evolution equations 
for the $g_\text{t}(t)$ and $g_\text{$\xi$}(\xi)$:
\begin{align}
\partial_t g_\text{t}(t) 
=
 - k_\text{t}(t) g_\text{t}(t),
\label{eq-gt-Dyn}
\end{align}
and 
\begin{align}
\gamma \partial_{\xi} g_\text{$\xi$}(\xi)
=
 - k_\text{$\xi$}(\xi) g_\text{$\xi$}(\xi),
\label{eq-gx-Dyn}
\end{align}
which are supplemented with the initial conditions
$g_\text{t}(t)=1$ at $t=0$ 
and
$g_\text{$\xi$}(\xi)=1$ at $\xi=\xi_0$. 
In this way $g_\text{$\xi$}(\xi)$ and $g_\text{t}(t)$ capture the decrease
of probability in $\Pa(\xi,t)$ due to the mass-related
and the time-related dropout processes, respectively.
Due to the additive structure of the dropout rate,
each $g$ can do that separately from the other.

Inserting 
$\Pone(\xi,t)$ from eq.~(\ref{eq-Pone-MLP-Dyn2}) into the ansatz eq.~(\ref{eq-Pa-ansatz}),
and that  
into eq.~(\ref{eq-PIMF-Dyn}) then gives us the IMF for 
$\xi \geq \xi_0$ as
\begin{align}
	P_\text{IMF}(\xi) 
	=  
	\int_{0}^{\infty}  ( k_\text{t}(t') + k_\text{$\xi$}(\xi) ) 
		\, g_\text{t}(t') \, g_\text{$\xi$}(\xi) \, \hat{\delta}(\xi- \xi_0 - \gamma t') \, \text{d}t'.
\label{eq-PIMF-ktkx}
\end{align}

%%TABLE%%%%TABLE%%%%TABLE%%%%TABLE%%%%TABLE%%%%TABLE%%%%TABLE%%
\begin{table}
	\caption{\label{tab:var}Notation and meaning of all parameters.}
	\begin{tabular}{ll}
	\hline
	Symbol & Meaning \\ \hline
	$\kappa_\text{t}$ 		& asymptotic maximum time-related dropout rate\\
	$\kappa_\text{$\xi$}$ 	& asymptotic maximum mass-related dropout rate\\
	$t_\text{St}$ 			& time-related Switch time,\\ 
						& when time-related dropout becomes large \\
	$\xi_\text{S$\xi$}$ 		& (logarithm of) mass-related Switch mass,\\ 
						& when mass-related dropout becomes large \\
	$1/\lambda_\text{t}$ 		& time span of the transition zone,\\
						& during which time-related dropout becomes large \\
	$1/\beta_\text{$\xi$}$ 	& mass span of the transition zone \\
						& during which mass-related dropout becomes large \\
	$m_0, \xi_0$ 			& initial mass, (logarithm of initial mass)  \\
	$t_0$ 				& initial time, onset of mass accretion, chosen to be 0 \\
	$\gamma$ 			& accretion rate \\
	$t_\text{m}$ 			& needed accretion time to grow $m_0$ to $m$ \\
	$m_\text{St}$ 			& time-related Switch mass,\\ 
						& when time-related dropout becomes large \\
	$m_\text{S$\xi$}$ 		& mass-related Switch mass,\\ 
						& when mass-related dropout becomes large\\
	$\alpha_\text{t}$ 		& dimensionless parameter, maximum time-related \\
						& dropout rate in terms of the accretion rate \\
	$\alpha_\text{$\xi$}$ 		& dimensionless parameter, maximum mass-related \\
						& dropout rate in terms of the accretion rate \\
	$\beta_\text{t}$ 			& dimensionless parameter, time span of the \\
						& transition zone in terms of the inv. accretion rate \\
	$\Pa$ 				& probability distribution of condensations (stars) \\
						& which actively accrete \\
	$\Ps$ 				& probability distribution of condensations (stars) \\
						& which stopped accreting \\
	$\Pone$ 				& probability distribution of condensations (stars) \\
						& which all always accrete due to missing dropout \\
	\hline
	\end{tabular}
\end{table}
%%TABLE%%%%TABLE%%%%TABLE%%%%TABLE%%%%TABLE%%%%TABLE%%%%TABLE%%

%SUBSECTION%%SUBSECTION%%SUBSECTION%%SUBSECTION%%SUBSECTION%
\subsection{The Truncated Multi Power Law IMF}

Now the IMF, as given by the above equation, can be evaluated in closed form
making use of the delta function features:
\begin{align}
P_\text{IMF}&(\xi) \;\text{d} (\xi)
\nonumber
\\
	&= \int_{0}^{\infty}  ( k_\text{t}(t') + k_\text{$\xi$}(\xi) ) 
		\, g_\text{t}(t') \, g_\text{$\xi$}(\xi) \, 
\\
		&\hspace{12ex}\times \hat{\delta}(\xi- \xi_0 - \gamma t') \, \text{d}t'
		\;\text{d} (\xi)
\nonumber
\\
	&= \int_{0}^{\infty}  ( k_\text{t}(t''/\gamma) + k_\text{$\xi$}(\xi) ) 
		\, g_\text{t}(t''/\gamma) \, g_\text{$\xi$}(\xi) \, 
\nonumber
\\
	&\hspace{12ex}\times \hat{\delta}(\xi- \xi_0 - t'') \, \text{d}t''/\gamma
		\;\text{d} (\xi)
\nonumber
\\
	&= ( k_\text{t}((\xi- \xi_0)/\gamma) + k_\text{$\xi$}(\xi) ) 
		\, g_\text{t}((\xi- \xi_0)/\gamma)\, 	
\nonumber
\\
	&\hspace{12ex}\times g_\text{$\xi$}(\xi)/\gamma
		\;\text{d} (\xi)
\nonumber 
\\
	&= \begin{cases}
		0                                                                                &:  \;  \xi < \xi_0  \\
		\frac{ (k_\text{t}(t_\text{m}) +  k_\text{$\xi$}(\xi) ) \,
		g_\text{t}(t_\text{m})\,g_\text{$\xi$}(\xi)}{\gamma}   \;\;\;\text{d} (\xi)   &:  \;  \xi \geq \xi_0,
	\end{cases}
\label{eq-PIMF-Delta-1}
\end{align}
where $t_\text{m} = (\xi- \xi_0)/{\gamma}$.
Inserting eq. (\ref{eq-kt-Def})
into 
eq. (\ref{eq-gt-Dyn})
and
eq. (\ref{eq-kxi-Def})
into
eq. (\ref{eq-gx-Dyn}),
while using the above initial conditions,
we obtain the solutions
\begin{align}
g_\text{t}(t)
&=
e^{-\frac{\kappa_\text{t}}{2} t} 
	\cosh[(t-t_\text{St})\lambda_\text{t}]^{-\frac{\kappa_\text{t}}{2 \lambda_\text{t}}}   
		\cosh[-t_\text{St} \lambda_\text{t}]^{\frac{\kappa_\text{t}}{2 \lambda_\text{t}}} ,
\label{eq-gt-Def} 
\\
g_\text{$\xi$}(\xi)
&=
e^{-\frac{\kappa_\text{$\xi$}}{2 \gamma} (\xi - \xi_0)} 
	\cosh[(\xi-\xi_\text{S$\xi$})\beta_\text{$\xi$}]^{-\frac{\kappa_\text{$\xi$}}{2 \gamma \beta_\text{$\xi$}}}   
\nonumber
\\	
	&\hspace{12ex}\times 
	\cosh[ (\xi_0 - \xi_\text{S$\xi$} )\beta_\text{$\xi$}]
		^{\frac{\kappa_\text{$\xi$}}{2 \gamma \beta_\text{$\xi$}}} .
\label{eq-gx-Def} 
\end{align}

\begin{figure}
      \includegraphics[width=\columnwidth]{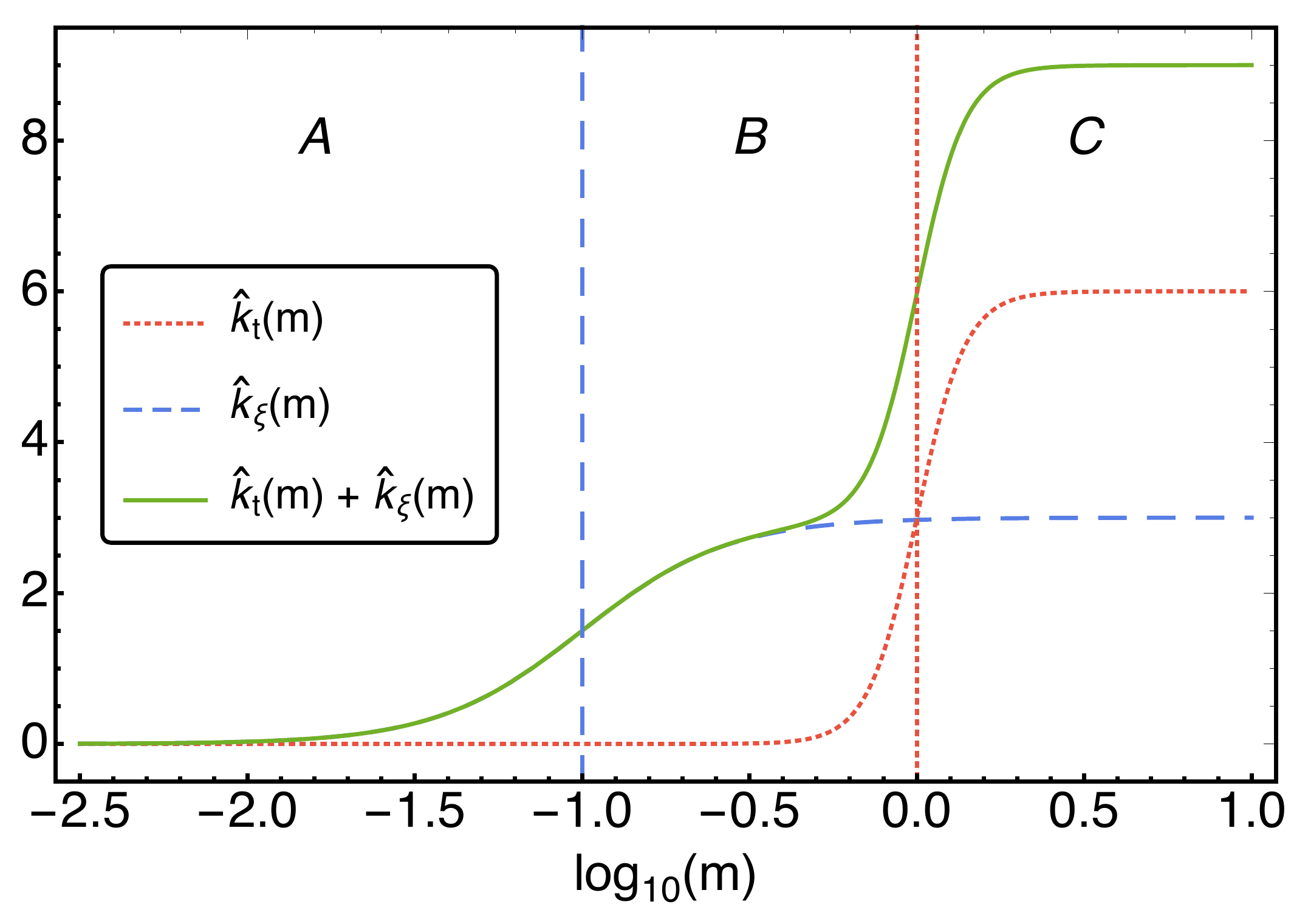}
      \caption{
      {The dropout rates $\hat k_\text{t}(m)$  
      and  $\hat k_\text{$\xi$}(m)$ given in eq.~(\ref{eq-HATkt-Def}) and (\ref{eq-HATkxi-Def}) 
      and their sum are given for the parameter set 
      $ \alpha_\text{t} = 6 $,
	$ \beta_\text{t} = 3 $,
	$ m_\text{St} = 10^{0} $,
	$ \alpha_\text{$\xi$} = 3 $,
	$ \beta_\text{$\xi$} = 1 $, and
	$ m_\text{S$\xi$}  = 10^{-1} $.
	We observe three regions: A low masses, B intermediate masses, and C high masses.}
      }
      \label{fig-LM-kt-kx-kt+kx}
    \end{figure}

We now introduce the 
time-related dropout switch mass 
$m_\text{St}$
via $ t_\text{St} = (\ln m_\text{St}- \ln m_0)/{\gamma}$
and
define dimensionless parameters
$\alpha_\text{t} = \frac{\kappa_\text{t}}{\gamma}$ and
$\beta_\text{t} = \frac{\lambda_\text{t}}{\gamma }$.
Then by substituting $t_\text{m}$ and $ t_\text{St} $ we find
\begin{align}
\hat g_\text{t}&(m,m_0)
=
\nonumber
\\	
&e^{-\frac{\alpha_\text{t}}{2} \ln \frac{m}{m_0}} 
	\cosh \left[ \beta_\text{t} \ln \frac{m}{m_\text{St}} \right]^{-\frac{\alpha_\text{t}}{2 \beta_\text{t}}}   
		\cosh \left[ \beta_\text{t} \ln \frac{m_0}{m_\text{St}} \right]^{\frac{\alpha_\text{t}}{2 \beta_\text{t}}}.
\label{eq-HATgt0-Def} 
\end{align}
Similarly we introduce the dimensionless parameter
$\alpha_\text{$\xi$} = \frac{\kappa_\text{$\xi$}}{\gamma}$
to obtain
\begin{align}
\hat g_\text{$\xi$}&(m,m_0)
=
\nonumber
\\	
&e^{-\frac{\alpha_\text{$\xi$}}{2} \ln \frac{m}{m_0}} 
	\cosh \left[ \beta_\text{$\xi$} \ln \frac{m}{m_\text{S$\xi$}} \right]^{-\frac{\alpha_\text{$\xi$}}{2 \beta_\text{$\xi$}}}   
		\cosh \left[ \beta_\text{$\xi$} \ln \frac{m_0}{m_\text{S$\xi$}} \right]^{\frac{\alpha_\text{$\xi$}}{2 \beta_\text{$\xi$}}} .
\label{eq-HATgx0-Def} 
\end{align}
In addition we rewrite the dropout rates as
\begin{align}
	\hat k_\text{t}(m) &= \frac{\alpha_\text{t}}{2} \left(1+\tanh \left[ \beta_\text{t} \ln \frac{m}{m_\text{St}} \right]  \right),
\label{eq-HATkt-Def} 
\\
	\hat k_\text{$\xi$}(m) &= \frac{\alpha_\text{$\xi$}}{2} \left(1+\tanh \left[ \beta_\text{$\xi$} \ln \frac{m}{m_\text{S$\xi$}} \right]  \right).
\label{eq-HATkxi-Def} 
\end{align}
To give a visualization of the dropout rates 
we plotted both, $\hat k_\text{t}(m)$ and $\hat k_\text{$\xi$}(m)$,
in figure \ref{fig-LM-kt-kx-kt+kx}
for the parameter set 
	$ \alpha_\text{t} = 6 $,
	$ \beta_\text{t} = 3 $,
	$ m_\text{St} = 10^{0} $,
	$ \alpha_\text{$\xi$} = 3 $,
	$ \beta_\text{$\xi$} = 1 $, and
	$ m_\text{S$\xi$}  = 10^{-1} $
also used later.

%FIGURE%FIGURE%FIGURE%FIGURE%FIGURE%FIGURE%FIGURE%FIGURE%FIGURE%FIGURE
%FIGURE%FIGURE%FIGURE%FIGURE%FIGURE%FIGURE%FIGURE%FIGURE%FIGURE%FIGURE

Combining these results we find for $m \geq m_0$
the truncated multi power law distribution $\Ptddpl$ as
\begin{align}
\Ptddpl&(m; \alpha_\text{t}, \beta_\text{t}, m_\text{St},\alpha_\text{$\xi$},\beta_\text{$\xi$}, m_\text{S$\xi$},m_0)  \;\text{d} (\ln m) 
\nonumber
\\
&= (  \hat k_\text{t}(m) + \hat k_\text{$\xi$}(m) )  \hat g_\text{t}(m,m_0) \hat g_\text{$\xi$}(m,m_0)\,\text{d} (\ln m).
\label{eq-Ptddpl-Def} 
\end{align}
We named this distribution the \emph{truncated multi power law distribution}
as it is defined only for $m \geq m_0$,
contrary to the \emph{multi power law distribution}
which will be defined below for all $m$.
The truncated multi power law distribution $\Ptddpl$
depends on seven dimensionless parameters.
$\alpha_\text{t} = \frac{\kappa_\text{t}}{\gamma}$ and
$\alpha_\text{$\xi$} = \frac{\kappa_\text{$\xi$}}{\gamma}$
measure the maximum dropout rates in terms of the
mass growth rate, which provides the characteristic time unit.
$\beta_\text{t} = \frac{\lambda_\text{t}}{\gamma }$ and
$\beta_\text{$\xi$}$
give the mass span of the transition zones of the
time- and mass-related dropout rates
in terms of $\ln m$.
Likewise $m_\text{St}$ and $m_\text{S$\xi$}$
define the switch masses for the onset of the increased dropout,
and finally $m_0$ the initial mass at the accretion onset.

%SECTION%%SECTION%%SECTION%%SECTION%%SECTION%%SECTION%%SECTION%
%SECTION%%SECTION%%SECTION%%SECTION%%SECTION%%SECTION%%SECTION%
\section{The Multi Power Law IMF}

In order to simplify the IMF further we note that we can 
take the limit  $m_0 \to 0$ 
and obtain as a limiting distribution the
Multi Power Law (MPL) IMF defined for all $m>0$:
\begin{align}
\Pddpl&(m; \alpha_\text{t}, \beta_\text{t}, m_\text{St},\alpha_\text{$\xi$},\beta_\text{$\xi$}, m_\text{S$\xi$})  \;\text{d}m
\nonumber
\\
&= (  \hat k_\text{t}(m) + \hat k_\text{$\xi$}(m) )  \;  \hat g_\text{t}(m)  \;  \hat g_\text{$\xi$}(m) / m \,\text{d}m 
\label{eq-Pddpl-Def}
\\
&= \hat k_\text{t}(m) \; \hat g_\text{t}(m)  \;  \hat g_\text{$\xi$}(m) 
	/ m \,\text{d}m
	+ \hat k_\text{$\xi$}(m) \; \hat g_\text{t}(m)  \;  \hat g_\text{$\xi$}(m)
	/ m \,\text{d}m 
\nonumber
\\
	&= \Pddplt + \Pddplxi ,
\nonumber 
\end{align}
with
\begin{align}
\hat g_\text{t}(m)
&=
2^{-\frac{\alpha_\text{t}}{2 \beta_\text{t}}} 
e^{-\frac{\alpha_\text{t}}{2} \ln \frac{m}{m_\text{St}}} 
	\cosh \left[ \beta_\text{t} \ln \frac{m}{m_\text{St}} \right]^{-\frac{\alpha_\text{t}}{2 \beta_\text{t}}}   \;,
\label{eq-HATgt-Def} 
\\
\hat g_\text{$\xi$}(m)
&=
2^{-\frac{\alpha_\text{$\xi$}}{2 \beta_\text{$\xi$}}} 
e^{-\frac{\alpha_\text{$\xi$}}{2} \ln \frac{m}{m_\text{S$\xi$}}} 
	\cosh \left[ \beta_\text{$\xi$} \ln \frac{m}{m_\text{S$\xi$}} \right]^{-\frac{\alpha_\text{$\xi$}}{2 \beta_\text{$\xi$}}}  \;,
\label{eq-HATgxi-Def} 
\end{align}
and $\hat k_\text{t}(m)$ and $\hat k_\text{$\xi$}(m)$ from eq. (\ref{eq-HATkt-Def}) and eq. (\ref{eq-HATkxi-Def}), respectively.

The transition from eq. (\ref{eq-HATgt0-Def}) and eq. (\ref{eq-HATgx0-Def}) to
eq. (\ref{eq-HATgt-Def}) and eq. (\ref{eq-HATgxi-Def}) is facilitated by
using
the small $x$ behavior of $\cosh[a x]  = (e^{-a x}+e^{a x})/2$.
For $a x \ll 0$ the first exponential dominates and we find
$\cosh[a x] \approx e^{-a x}/2$.
Applying this on the $m_0-\cosh$-term in eq. (\ref{eq-HATgt0-Def})
for $a=\beta_\text{t}>0$ and $m_0 \ll m_\text{St}$ and thus $x=\ln \frac{m_0}{ m_\text{St} }\ll 0$
one gets
\begin{align}
\hat g_\text{t}&(m,m_0) 
\nonumber
\\
&\approx
e^{-\frac{\alpha_\text{t}}{2} \ln \frac{m}{m_0}} 
	\cosh \left[ \beta_\text{t} \ln \frac{m}{m_\text{St}} \right]^{-\frac{\alpha_\text{t}}{2 \beta_\text{t}}}   
		2^{-\frac{\alpha_\text{t}}{2 \beta_\text{t}}} 
\left(e^{ - \beta_\text{t} \ln \frac{m_0}{m_\text{St}}}\right)^{ \frac{\alpha_\text{t}}{2 \beta_\text{t}}} 
\nonumber
\\
&=e^{-\frac{\alpha_\text{t}}{2} \ln \frac{m}{m_0}} 
	\cosh \left[ \beta_\text{t} \ln \frac{m}{m_\text{St}} \right]^{-\frac{\alpha_\text{t}}{2 \beta_\text{t}}}   
		2^{-\frac{\alpha_\text{t}}{2 \beta_\text{t}}} 
e^{ -\frac{\alpha_\text{t}}{2} \ln \frac{m_0}{m_\text{St}}}
\nonumber
\\
&=
2^{  -\frac{  \alpha_\text{t}  }{   2 \beta_\text{t}  }   }
e^{ -\frac{  \alpha_\text{t}  }{  2  }  \ln \frac{m}{m_\text{St}}  }
\cosh \left[ \beta_\text{t} \ln \frac{m}{m_\text{St}} \right]^{-\frac{\alpha_\text{t}}{2 \beta_\text{t}}}  ,
\label{eq-HATgt0-Def-smallM0} 
\end{align}
which shows, that the leading term of this approximation is no longer $m_0$-dependent.

Applying in addition the above reasoning to the remaining $\cosh$-term
in $\hat g_\text{t}(m) $
for $a=\beta_\text{t}>0$ and $m \gg m_\text{St}$ and thus $x=\ln \frac{m}{ m_\text{St} }\gg 0$
leads to the large $m$ behavior of  
\begin{align}
\hat g_\text{t}&(m) 
\nonumber
\\
&\approx
2^{  -\frac{  \alpha_\text{t}  }{   2 \beta_\text{t}  }   }
e^{ -\frac{  \alpha_\text{t}  }{  2  }  \ln \frac{m}{m_\text{St}}  }
2^{  \frac{  \alpha_\text{t}  }{   2 \beta_\text{t}  }   }
\left( e^{\beta_\text{t} \ln \frac{m}{m_\text{St}}} \right)^{-\frac{\alpha_\text{t}}{2 \beta_\text{t}}}  
\nonumber
\\
&=
\left(\frac{m}{m_\text{St}} \right)^{ -\frac{  \alpha_\text{t}  }{  2  }   }
e^{ -\frac{\alpha_\text{t}}{2} \ln \frac{m}{m_\text{St}}}
=
\left(\frac{m}{m_\text{St}} \right)^{ - \alpha_\text{t}  },
\label{eq-cosh-simp-7} 
\end{align}
showing a power law decay for $\hat g_\text{t}(m)$. 
Similarly, $\hat g_\text{$\xi$}(m,m_0)$ can then be handled
and leads to $\hat g_\text{$\xi$}(m)$,
which also shows a power law.

There are a number of interesting features of the MPL,
some of which we will discuss in the following.
Note that from a technical perspective 
the functional form of the time-dependent dropout rate
and the mass-dependent dropout rate are equivalent.

The parameter set used as a standard set in the subsequent discussion is
$ \alpha_\text{t} = 6 $,
$ \beta_\text{t} = 3 $,
$ m_\text{St} = 10^{0} $,
$ \alpha_\text{$\xi$} = 3 $,
$ \beta_\text{$\xi$} = 1 $, and
$ m_\text{S$\xi$}  = 10^{-1} $.
These parameters are chosen exclusively for demonstrating the effects on the resulting IMF
and not for a direct astrophysical application.

%FIGURE%FIGURE%FIGURE%FIGURE%FIGURE%FIGURE%FIGURE%FIGURE%FIGURE%FIGURE
\begin{figure}
      \includegraphics[width=\columnwidth]{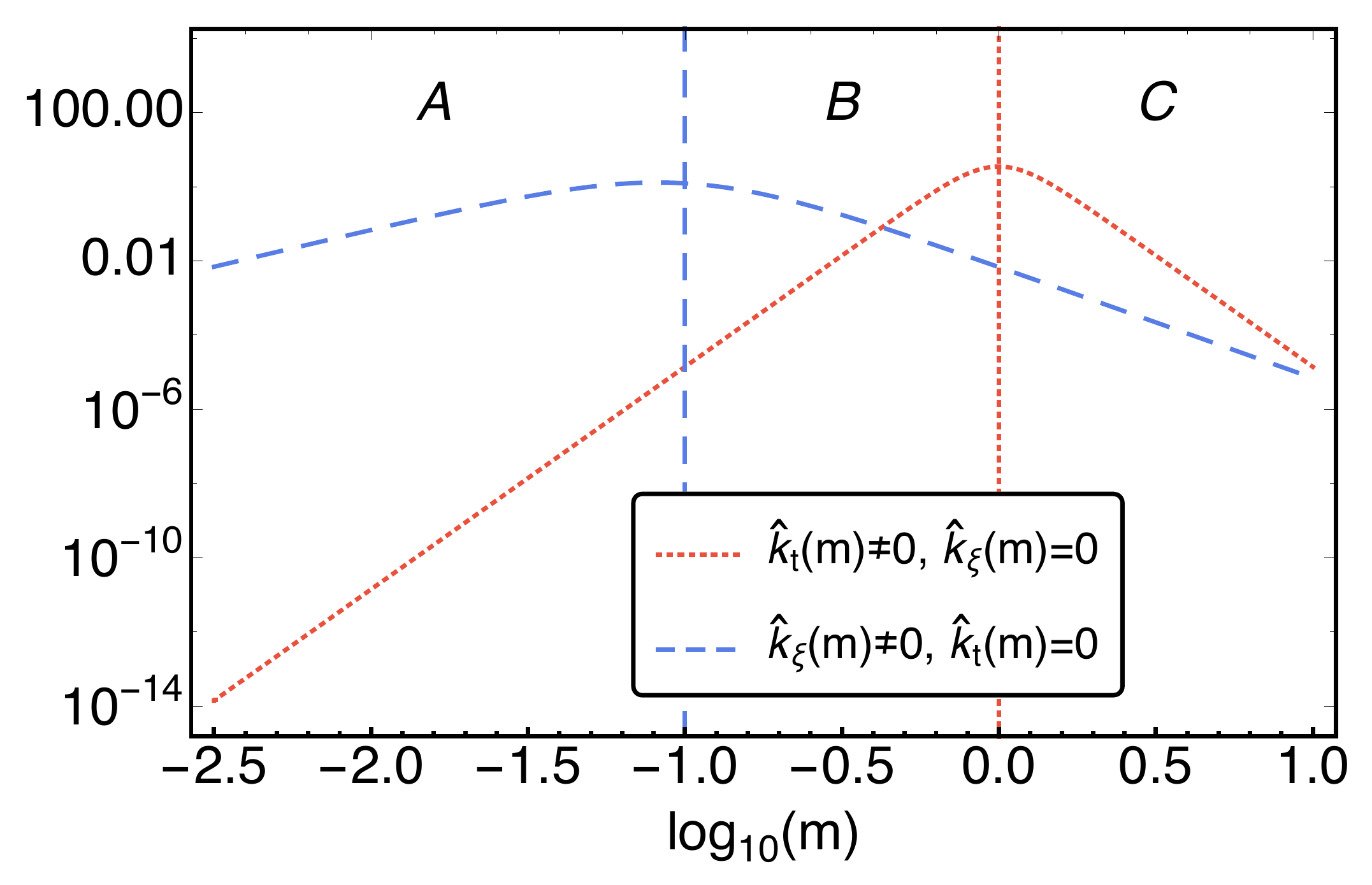}
      \caption{
      The MPL distribution given in eq.~(\ref{eq-Pddpl-Def}) 
      is shown for two cases: 
      one in which the mass dropout rate $\hat k_\xi$ is set to 0
      ($\alpha_\text{$\xi$} = 0$) 
      but the time dropout rate $\hat k_\text{t}$ is nonzero 
      ($\alpha_\text{t} = 6$,
		$\beta_\text{t} = 3 $, and
		$m_\text{St}  = 10^{0} $),
      and one in which the time dropout rate $\hat k_\text{t}$ is set to 0
      ($\alpha_\text{t} = 0$)
      but the mass dropout rate $\hat k_\xi$ is nonzero
      ($\alpha_\text{$\xi$} = 3 $,
		$\beta_\text{$\xi$} = 1 $, and
		$m_\text{S$\xi$} = 10^{-1} $). 
      Here the low and the high mass end 
      of the MPL distribution with $\hat k_\xi(m) =0$
      has a steeper increase and decrease, respectively, 
      than for the MPL distribution with $\hat k_\text{t}(m) = 0$.}
      \label{fig-MPLkt=0-MPLkx=0}
    \end{figure}
%FIGURE%FIGURE%FIGURE%FIGURE%FIGURE%FIGURE%FIGURE%FIGURE%FIGURE%FIGURE
In figure
\ref{fig-MPLkt=0-MPLkx=0}
we show the 
$\Pddpl(m; \alpha_\text{t}, \beta_\text{t}, m_\text{St},\alpha_\text{$\xi$},\beta_\text{$\xi$}, m_\text{S$\xi$})$
for two variations of the standard set: 
one for which the mass dropout rate $\hat k_\xi$ is set to 0 
($ \alpha_\text{$\xi$} = 0 $)
but the time dropout rate $\hat k_\text{t}$ is nonzero
($ \alpha_\text{t} = 6$, 
$ \beta_\text{t} = 3 $, 
and $ m_\text{St} = 10^{0} $),
the other for which the time dropout rate $\hat k_\text{t}$ is set to 0 
($ \alpha_\text{t} = 0 $)
but the mass dropout rate $\hat k_\xi$ is nonzero
($ \alpha_\text{$\xi$} = 3 $, 
$ \beta_\text{$\xi$} = 1 $, and 
$ m_\text{S$\xi$} = 10^{-1} $).
In both cases one finds the familiar form of the DPL distribution.
We can immediately recognize three regions A, B, and C:
region A where both MPLs increase, 
region B where the MPL with $\hat k_\text{t}=0$ decays 
while the one with $\hat k_\text{$\xi$}=0$ still increases, and
region C where both decay.
In each region there is a clear power law behaviour 
appearing in this log-log plot as straight lines.

We notice that the decay of the MPLs for high masses is due to
the decay of $\hat g_\text{t}(m)$ and of $\hat g_\text{$\xi$}(m)$.
These functions decay with a power law.
Their respective exponents are $-\alpha_\text{t}$ and $-\alpha_\text{$\xi$}$.
The dropout rates $\hat k_\text{t}(m)$ and $\hat k_\text{$\xi$}(m)$
on the other hand increase with power laws 
with exponents $2 \beta_\text{t}$ and $2 \beta_\text{$\xi$}$, respectively.
The symmetry of the MPL with a mass dropout rate $\hat k_\xi= 0$ 
makes it easy
to recognize this: $ \alpha_\text{t} = 6  = 2 \beta_\text{t}$.

%FIGURE%FIGURE%FIGURE%FIGURE%FIGURE%FIGURE%FIGURE%FIGURE%FIGURE%FIGURE
\begin{figure}
      \includegraphics[width=\columnwidth]{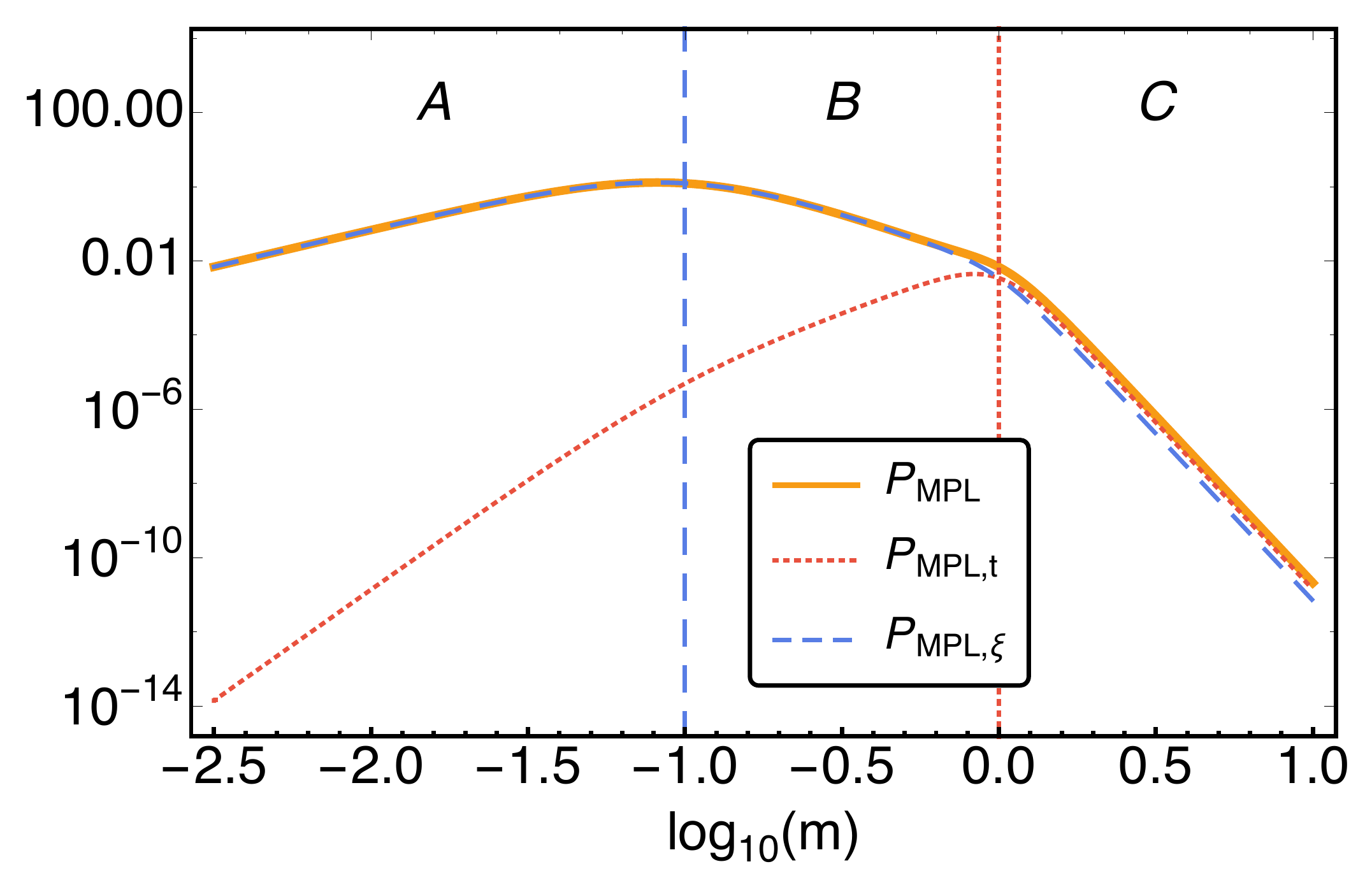}
      \caption{
      $\Pddpl$ and the corresponding time term $\Pddplt$ and mass term $\Pddplxi$, 
      as given in eq.~(\ref{eq-Pddpl-Def}), are shown for the standard parameter set 
      ($\alpha_\text{t} = 6 $,
		$\beta_\text{t} = 3 $,
		$m_\text{St} = 10^{0} $,
		$\alpha_\text{$\xi$} = 3 $,
		$\beta_\text{$\xi$} = 1 $, and
		$m_\text{S$\xi$}  = 10^{-1} $). 
      The overall MPL distribution shows three different regions. 
      In region A and B the mass term $\Pddplxi$
      and in C the time term $\Pddplt$ dominates the overall behaviour.
      }
      \label{fig-ktMPL-kxMPL}
    \end{figure}
%FIGURE%FIGURE%FIGURE%FIGURE%FIGURE%FIGURE%FIGURE%FIGURE%FIGURE%FIGURE

We now turn our attention to the MPL with the base parameter set and
investigate the contributions of the two dropout rates to the overall MPL.
These are shown in figure \ref{fig-ktMPL-kxMPL}.
Due to the multiplicative structure of the MPL with respect to 
$\hat g_\text{t}(m)$ and to $\hat g_\text{$\xi$}(m)$
the overall decay of the MPL for masses greater than
the maximum of $m_\text{St}$ and $m_\text{S$\xi$}$ (i.e. in region C)
is with a power law with exponent $\nu_> = -\alpha_\text{t}-\alpha_\text{$\xi$}-1$,
where the $-1$ is due to the Jacobian in the transition from $\log m$ to $m$.
It is important to note that this decay also applies to each of the two contributions separately.
In region A and B the mass dropout rate dominates the overall behaviour
and thus the MPL increases in A with a power law exponent $2 \beta_\text{$\xi$} - 1 $
and decreases in B with $ - \alpha_\text{$\xi$} - 1 $.
The time dropout contribution increases in A with $2 \beta_\text{t} - 1 $
and in B with $2 \beta_\text{t} - \alpha_\text{$\xi$} - 1 $.
Interestingly, the time dropout seems to have only a very minor influence on the MPL 
in regions A and B
but its presence shows up in the increased decay in region C.

\begin{figure}
      \includegraphics[width=\columnwidth]{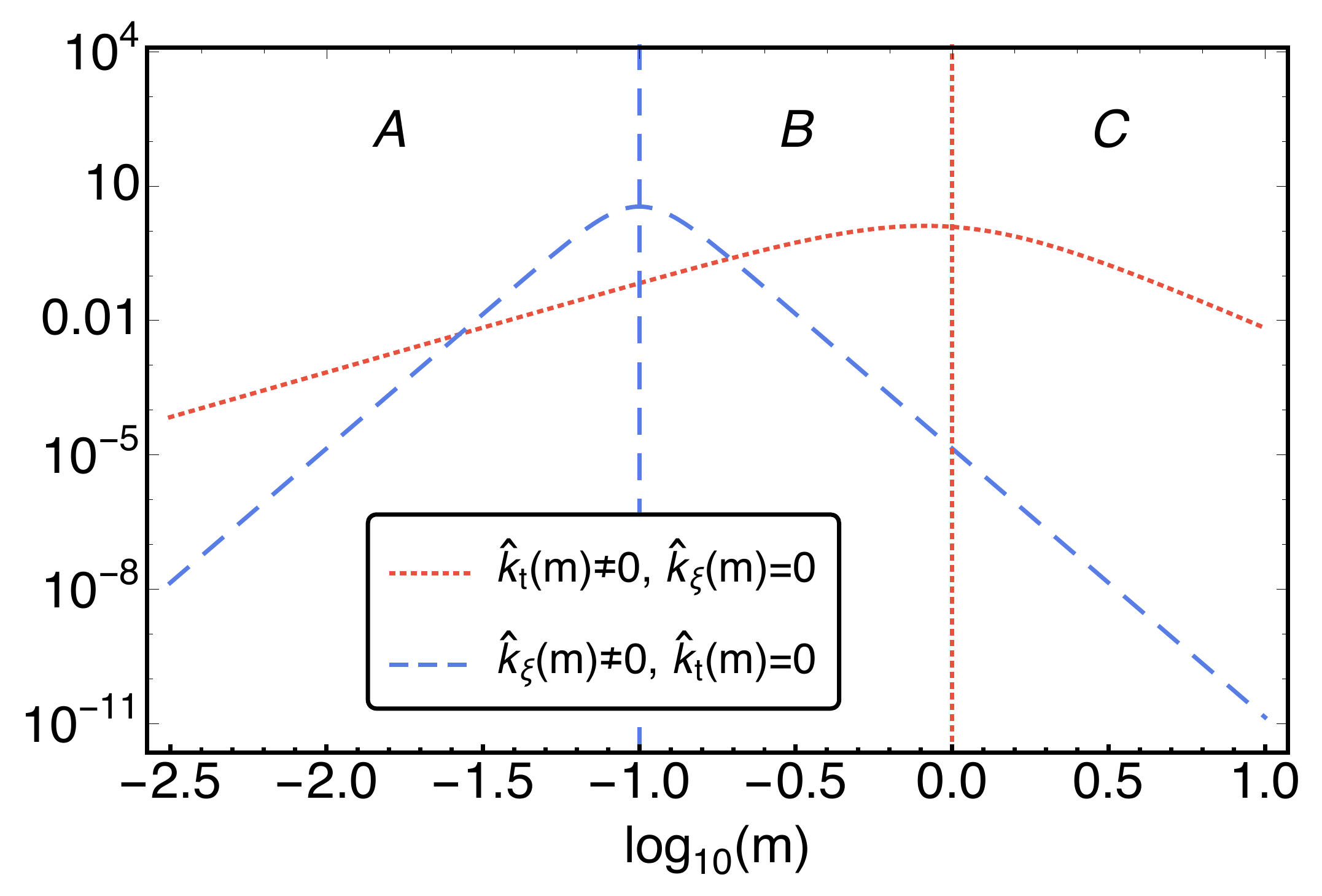}
      \caption{The MPL distribution given in eq.~(\ref{eq-Pddpl-Def}), 
      is shown for two cases: 
      one in which the mass dropout rate $\hat k_\xi$ is set to 0
      ($\alpha_\text{$\xi$} = 0 $) 
      but the time dropout rate $\hat k_\text{t}$ is nonzero
      ($\alpha_\text{t} = 3$
		$\beta_\text{t} = 1 $, and
		$m_\text{St}  = 10^{0} $)}
      and one in which the time dropout rate $\hat k_\text{t}$ is set to 0
      ($\alpha_\text{t} = 0 $)
      but the mass dropout rate $\hat k_\text{$\xi$}$ is nonzero
      {($\alpha_\text{$\xi$} = 6 $,
		$\beta_\text{$\xi$} = 3 $, and
		$m_\text{S$\xi$} = 10^{-1} $)}. 
      Here the low and the high mass end 
      of the MPL distribution with $\hat k_\text{t}(m) =0$
      has a steeper increase and decrease, respectively, 
      than for the MPL distribution with $\hat k_\xi(m) = 0$.
      \label{fig-MPLkt=0-MPLkx=0-invert}
    \end{figure}

The situation changes drastically if we make a change in the standard parameter set
by exchanging the values of $\alpha_\xi$ with $\alpha_\text{t}$
and $ \beta_\xi $ with $\beta_\text{t}$:
$\alpha_\text{t} = 3 $,
$\beta_\text{t} = 1 $,
$\alpha_\text{$\xi$} = 6 $,
$\beta_\text{$\xi$} = 3 $, 
and
Figure \ref{fig-MPLkt=0-MPLkx=0-invert} shows the MPL again for
the two cases of vanishing time and mass dropout.

%FIGURE%FIGURE%FIGURE%FIGURE%FIGURE%FIGURE%FIGURE%FIGURE%FIGURE%FIGURE
%FIGURE%FIGURE%FIGURE%FIGURE%FIGURE%FIGURE%FIGURE%FIGURE%FIGURE%FIGURE

\begin{figure}
      \includegraphics[width=\columnwidth]{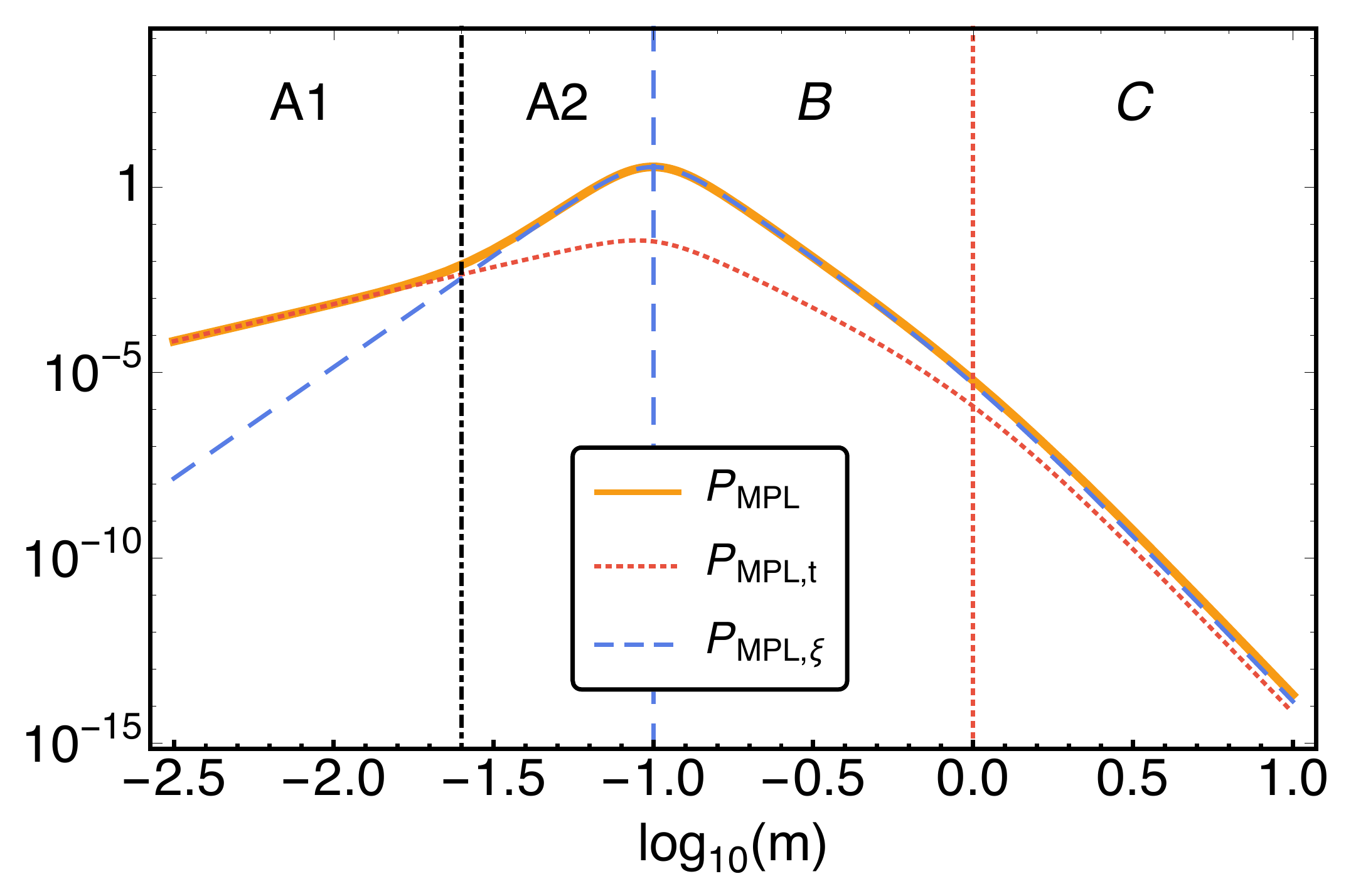}
      \caption{$\Pddpl$ and the corresponding time term $\Pddplt$ and mass term $\Pddplxi$, 
      as given in eq.~(\ref{eq-Pddpl-Def}), are
      shown for the altered standard parameter set 
      ($ \alpha_\text{t} = 3 $,
		$ \beta_\text{t} = 1 $,
		$ m_\text{St} = 10^{0} $,
		$ \alpha_\text{$\xi$} = 6 $,
		$ \beta_\text{$\xi$} = 3 $, and
		$ m_\text{S$\xi$}  = 10^{-1} $).
		As $\Pddpl$ is the sum of $\Pddplt$, and $\Pddplxi$, 
      here a complex behaviour of the MPL distribution,
      with four different regions is found.
      }
      \label{fig-ktMPL-kxMPL-invert}
    \end{figure}

If we now combine the two contributions, the MPL shows a distinctly different
form than in the standard case as can be seen in figure \ref{fig-ktMPL-kxMPL-invert}.
As the two contributions cross in region A 
and the mass rate becomes greater than the time rate contribution,
region A is split into regions A1 and A2.
In A1 the power law increase is with the smaller exponent $2 \beta_\text{t} -1 = 1 $,
while in A2 the mass dropout rate  with $2 \beta_\text{$\xi$} -1 = 3 $ takes over.
This is shown in figure \ref{fig-ktMPL-kxMPL-invert}.
In region B we find an exponent $ 2 \beta_\text{t} - \alpha_\text{$\xi$} -1$
and in C the exponent equals $ \alpha_\text{t} - \alpha_\text{$\xi$} -1$.

%FIGURE%FIGURE%FIGURE%FIGURE%FIGURE%FIGURE%FIGURE%FIGURE%FIGURE%FIGURE
%FIGURE%FIGURE%FIGURE%FIGURE%FIGURE%FIGURE%FIGURE%FIGURE%FIGURE%FIGURE

%SECTION%%SECTION%%SECTION%%SECTION%%SECTION%%SECTION%%SECTION%
%SECTION%%SECTION%%SECTION%%SECTION%%SECTION%%SECTION%%SECTION%
\section{Illustrative Example: Orion Nebula Cluster IMF}

In order to show the implications of our model 
let us assume that the dropout rate has a second increase
due to some global event on "Nebula time``,
i.e. it effects all stars in the same fashion
as shown in figure \ref{fig-LM-kt-kx-kt+kx}.
Of course Nebula time here is simply a modeling device introduced to illustrate 
this further increase in the accretion dropout rate,
which can in principle occur prior or after the basic mass-related dropout increase.
Such a further increase may be due to either
another intrinsic mass-related effect or 
to extrinsic (i.e. non single protostar mass-related) features of the nebula. 
For example, an extrinsic effect
can be that at a time $t_\text{St}$
after a molecular cloud starts forming,
the nebula may be largely cleared away.
Stars that formed early in the history of star formation in the molecular cloud
and happen to still be accreting
will then face a significant increase in the probability of stopping accretion.

Nebulae where active star formation takes place, 
are  firmly evolving according to common underlying physics. 
However star forming regions are each unique in the manner of complex systems. 
Each may develop in completely unique ways within physical law globally,  
and locally too.  
Local departures from universality might cancel out in the IMF, 
or they can be cumulative, leaving distinctive unique global features in the final PDF.

This presents a conundrum for an idealized notion of  the IMF 
that applies to all clusters producing a common result.
Each cluster has a unique history that limits the consideration of universal properties. 
Moreover, young clusters like the ONC have mass distributions that are still forming. 
Furthermore,  there is reason to believe that there is more 
than one distinct population in play 
\cite{jerabkova.t.19.when.06974v2}

The notion of nebula time is a step toward capturing 
more complex histories and transient features. 
But the nebula time machinery is far from comprehensive in this regard.  
It represents a modeling experiment to gain a sense of  
how the evolution equation approach can be made to capture more subtle structures. 
Its  machinery in this paper still represents a considerable constraint 
on what forms are possible. 
Is it flexible enough to match actual data of some kind?
To address this question 
we present an example for the range of possible IMF forms
which can be achieved within our modeling approach.
In particular we show that the MPL IMF can fit observed IMF data well. 
This does not constitute a claim 
that the data in question represents a system with a Nebula time 
and it does not represent an empirical fit of data. 
It merely establishes that physically coherent sub-classes of PDFs 
can effectively match sensible data, 
giving us clues for next steps of development in the evolution equation approach.

As a basis for this exercise we use data presented by 
\cite{rio.n.12.initial.14}
for the Orion Nebula Cluster
(here taken from fig.~18 of 
\cite{rio.n.12.initial.14}).
%FIGURE%FIGURE%FIGURE%FIGURE%FIGURE%FIGURE%FIGURE%FIGURE%FIGURE%FIGURE
\begin{figure}
      \includegraphics[width=\columnwidth]{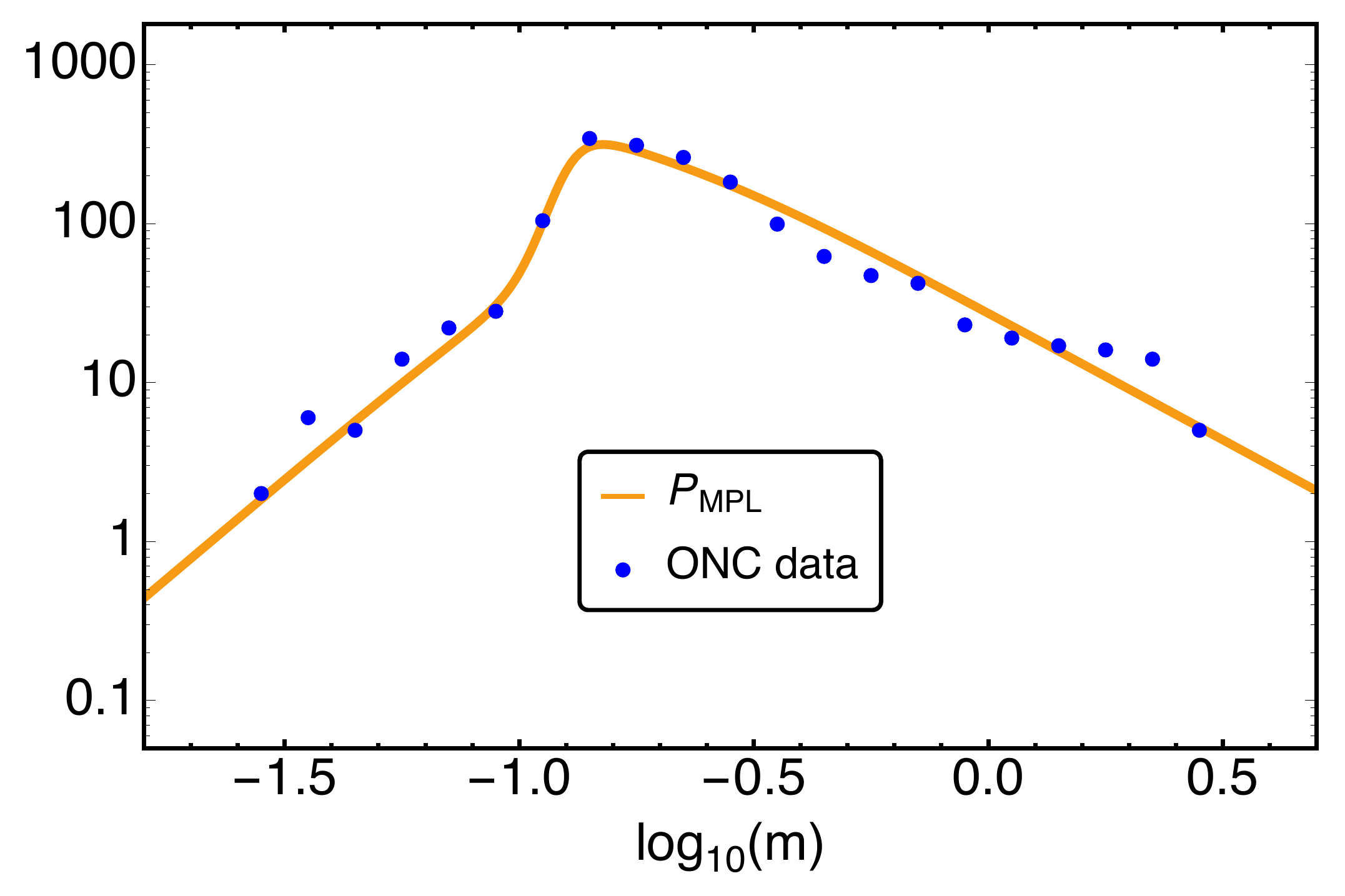}
       \caption{The IMF for the Orion Nebula Cluster as obtained by 
       \citep{rio.n.12.initial.14} 
       is shown 
       together with an appropriately adjusted MPL distribution.
       }
      \label{fig-MPL-ONC}
    \end{figure}
%FIGURE%FIGURE%FIGURE%FIGURE%FIGURE%FIGURE%FIGURE%FIGURE%FIGURE%FIGURE
Figure \ref{fig-MPL-ONC} shows a fit
for the ONC data of
\cite{rio.n.12.initial.14}
based on the D'Antona \& Mazzitelli model.
The parameters are 
$ \alpha_\text{t} = 0.6 $,
$ \beta_\text{t} = 1.25 $,
$ m_\text{St} = 10^{-0.72} $,
$ \alpha_\text{$\xi$} = 1.0 $,
$ \beta_\text{$\xi$} = 7.0 $, and
$ m_\text{S$\xi$}  = 10^{-0.9} $.
It follows that 
the exponent $\nu_> = -\alpha_\text{t}-\alpha_\text{$\xi$}-1= -2.6$
for the high mass power law decay is consistent with
\cite{
kroupa.p.02.initial.82}.
An interesting observation is that the two dropout masses are only a factor of $10^{0.2} \approx 1.6$
apart.

%SECTION%%SECTION%%SECTION%%SECTION%%SECTION%%SECTION%%SECTION%
%SECTION%%SECTION%%SECTION%%SECTION%%SECTION%%SECTION%%SECTION%
\section{Conclusions}

In this model we studied star formation within a dynamical picture based on evolution
equations for the probability distribution functions of protostars (still-accreting stars) of a particular mass and of stars that have dropped out of the accretion process. 
In particular we investigated the properties of IMFs that are generated if the dropout rate
increases in two (smeared out) steps. 
The resulting multi power law IMF is characterized
by a power law behaviour for low and high masses, where the respective exponents can
be determined in terms of the dropout rates.
In the intermediate regime, further power laws with different exponents emerged.
For instance, IMFs with three distinct sections can be generated and resemble 
forms that have been used before in the literature.
In this model the dropout rate consists of two parts, one of which describes accretion termination
based on star mass, and one which describes accretion termination
based on the evolution time; we refer to the latter as a nebula time
and it may be associated with processes that starve mass accretion such 
as outflows or feedback from other stars. 
The mass growth law is restricted to be exponential, 
and may be most applicable to late time 
accretion that leads to intermediate and high mass stars. 
Future work can explore the low mass accretion rate more realistically.
It could also include episodic events of mass growth.

The progress reached in this work lies in the demonstration of the variability
of IMFs possible within our approach. Although power laws are generated, the 
details of the value of particularly the nebula time may vary from one star-forming
cluster to another. Hence the physics of star formation may lead to power-laws
in the IMF, but leave behind a variety of index values and transition points. 
A compendium of fits to cluster mass functions reveals a noticeable but not
dramatic spread of power laws and transition points
\citep{bastian.n.10.universal.339}, 
but a debate continues as to whether or not this represents universality given
the observational uncertainties. Other theoretical models of the IMF such 
as turbulent fragmentation 
\citep{padoan.p.02.stellar.870,hennebelle.p.08.analytical.395} 
and ideas such as competitive accretion 
\citep{bonnell.i.97.accretion.201} 
rely on specific conditions of molecular
gas clouds and turbulence that need to be universal if the IMF is universal, 
and their dependence on varying initial gas conditions 
is difficult to model analytically.
Our approach is consistent with variability in a manner that may be quantifiable
in future work that ties the mass growth law and termination probabilities more 
closely to physical phenomena. It is also flexible enough to explain multiple
mass regions of the IMF and not just the intermediate to high mass tail.

%SECTION%%SECTION%%SECTION%%SECTION%%SECTION%%SECTION%%SECTION%
%SECTION%%SECTION%%SECTION%%SECTION%%SECTION%%SECTION%%SECTION%
\section{Acknowledgements}
We thank the referee Hans Zinnecker for his extremely thoughtful comments
that helped to improve the manuscript.

%%%%%%%%%%%%%%%%%%%%%%%%%%%%%%%%%%%%%%%%%%%%%%%%%%



%%\bibliographystyle{mnras}
%%\bibliography{ms_mnras.R2} % if your bibtex file is called example.bib
%%

% % Alternatively you could enter them by hand, like this:
% % This method is tedious and prone to error if you have lots of references
% \begin{thebibliography}{99}
% \bibitem[\protect\citeauthoryear{Author}{2012}]{Author2012}
% Author A.~N., 2013, Journal of Improbable Astronomy, 1, 1
% \bibitem[\protect\citeauthoryear{Others}{2013}]{Others2013}
% Others S., 2012, Journal of Interesting Stuff, 17, 198
% \end{thebibliography}

%%%%%%%%%%%%%%%%%%%%%%%%%%%%%%%%%%%%%%%%%%%%%%%%%%

% %%%%%%%%%%%%%%%%% APPENDICES %%%%%%%%%%%%%%%%%%%%%
% 
% \appendix
% 
% \section{Some extra material}
% 
% If you want to present additional material which would interrupt the flow of the main paper,
% it can be placed in an Appendix which appears after the list of references.
% 
% %%%%%%%%%%%%%%%%%%%%%%%%%%%%%%%%%%%%%%%%%%%%%%%%%%

% Don't change these lines
\bsp	% typesetting comment
\label{lastpage}
\end{document}